\documentclass[aps,twocolumn,amsmath,amssymb,floatfix,longbibliography]{revtex4-1}

\usepackage{txfonts}
\usepackage{microtype}
\usepackage{graphicx}
\usepackage{color}
\usepackage{ulem}
\usepackage{bm}
\usepackage[english]{babel}
\usepackage{tensind}
\usepackage{mathtools}
\usepackage[a4paper, total={7in, 10in}]{geometry}

\newcommand{\comm}[1]{}

\newcommand{\Tr}{{\rm Tr}\,}

\begin{document}

\title{\bf{Nonspreading relativistic  electron wavepacket in a strong laser field}}

\author{Andre G. Campos}
\email{agontijo@mpi-hd.mpg.de}
\author{Karen Z. Hatsagortsyan}
\author{Christoph H. Keitel}
\affiliation{Max Planck Institute for Nuclear Physics, Heidelberg 69117, Germany}

\date{\today}

\begin{abstract}

A solution of the Dirac equation in a strong laser field presenting a nonspreading wave packet \textit{in the rest frame} of the electron is derived. It consists of a generalization of the self-accelerating free electron wave packet [Kaminer \textit{et al.} Nature Phys. 11, 261 (2015)] to the case with the background of a strong laser field. Built upon the notion of nonspreading for an extended relativistic wavepacket, the concept of Born rigidity for accelerated motion in relativity is the key ingredient of the solution. At its core, the solution comes from the connection between the self-accelerated free electron wave packet and the eigenstate of a Dirac electron in a constant and homogeneous gravitational field via the equivalence principle. The solution is an essential step towards the realization of the laser-driven relativistic collider [Meuren \textit{et al.} PRL 114, 143201 (2015)], where the large spreading of a common Gaussian wave packet during the excursion in a strong laser field strongly limits the expectable yields.
 
\end{abstract}

\maketitle

\textit{Introduction.} Recent advances in ultrastrong laser technology  \cite{Danson_2019, Radier_2022,Yoon_2021} provide bright prospects for laser-driven particle acceleration techniques. Especially successful are laser-driven plasma-based accelerators \cite{Esarey_2009}, which raised hopes to develop further the technique to compete with conventional electron-positron colliders \cite{Leemans_2009, RMP_2012}, reducing the scale of the accelerating device. Even more dramatic scale change promises the idea of the laser-driven coherent microscopic collider \cite{Henrich_2004,Hatsagortsyan_2006,Muller_2008a,Meuren_2015},
where the electron and positron generation, acceleration, and collision are realized within a single stage in a microscopic scale, providing high luminosity due to the coherently controlled electron-positron recollision. The bottleneck of this idea is the large spreading of a single electron wave packet in the rest frame of the electron during the excursion in the laser field within one laser period, which significantly restrains the luminosity of the collision. Thus, the covet is the overriding of the wavepacket spreading for the electron motion in the continuum. Nonspreading free electron wave packets via interference of different momentum components in the wavepacket, so-called particle Airy beams, are known for the Schr\"odinger equation \cite{Berry_1979,Voloch-Bloch_2013}, which generalize the similar idea for optical beams  \cite{Siviloglou_2007,Baumgartl_2008,Chong_2010,Kaminer_2011,kaminer2012nondiffracting}. However, Airy beams are not normalizable, i.e., span the whole space. Because of the infinite extension of such wavepackets in space, they are not applicable for a laser-driven collider, as the luminosity of the collider should be quenched.

In the nonspreading wavepacket, the distance between two points remains constant during the motion. While the latter has a well-defined meaning in nonrelativistic mechanics, in the relativistic case, surprises arise, particularly involving Bell's paradox~\cite{Dewan_1959}. In this \textit{Gedankenexperiment} two points connected by a thread move with a constant acceleration keeping a constant distance between them in the Lab-frame, however, the thread between the points is broken because of the contracted length of the thread in the Lab-frame \cite{marzlin2014interferometry}. Then, how do the two points have to move to avoid breaking the thread connecting them? This question is resolved by the Born rigidity concept~\cite{Born_1909}, defining the notion of a rigid body in a relativistic setting: The wordlines of the rigid body points have to be equidistant curves in spacetime. Or in more simple terms, the space distance between two infinitesimally close points measured simultaneously in the co-moving inertial frame (rest frame) should be constant. In particular this will be the case, and the thread will not break in Bell's paradox, if the points move with different constant accelerations along hyperbolic trajectories \cite{Moeller}. In the Lab-frame the space distance between the infinitesimally close points will decrease, fitting to the Lorentz contraction, while the distance between them in the rest frame will remain constant. Note that for the luminosity of the laser-driven collider, namely the rest frame size of the electron and positron wave packets  matters at the recollision.

Although seemingly unrelated, generating nonspreading wavepackets in relativistic quantum mechanics shares a common thread with the resolution of Bell's paradox through the concept of Born rigidity and hyperbolic motion. In both cases, the motion of different points of the objects is crucial, whether it is the motion of the points of the rigid body
or the dynamics of interference fringes of the electron wavepacket along hyperbolic trajectories in the case of quantum mechanics. 

In this Letter, inspired by the geometrical concept of Born rigidity,  we use the Covariant Relativistic Dynamical Inversion (CRDI) technique \cite{Campos_2022} to demonstrate the existence of nonspreading wavepackets in a laser field fulfilling the  Born rigidity requirements. These wavepackets in the local rest frame of the electron feature interference fringes with a constant distance between them due to the fringes' dynamics along the hyperbolic trajectories \cite{Kaminer_2015}. Employing the CRDI technique, we develop a procedure to transform the wavepacket in the laser field to the local rest frame of the electron, where it evolves into a free electron wavepacket. To impose nonspreading property on the wavepacket fringes, we invoke the equivalence principle, which tells us that the hyperbolic trajectories, i.e., trajectories corresponding to a motion with constant acceleration are similar to those in a constant gravitational field. The latter allows us the construction of the nonspreading free electron wavepacket via mimicking locally the exact solution of the Dirac equation for the electron in a constant and homogeneous gravitational field \cite{Greiner_1985}. We have identified the finite lifetime of the nonspreading wavepacket because of the leaking from the Rindler space and proved that it is sufficient to allow recollision in a laser-driven collider.

\textit{Born's rigidity} Our main aim is to create relativistic nonspreading wavepackets in a sense that the distance between the wavepacket’s fringes remains constant with time in the electron’s local rest frame. The Born rigidity concept tells us that this aim will be realized if the dynamics of fringes of the wavepacket manifests hyperbolic trajectories along the so-called \textit{Rindler coordinates \cite{Moeller}:} $X=x,\, Y=y,\, Z=\frac{1}{g}\left(\cosh(gt)-1\right)+z\cosh(gt),\, 
T=\frac{1}{g}\sinh(gt) +z\sinh(gt)$, with the coordinates $(T,X,Y,Z)$ and $(t,x,y,z)$ in the Lab- and co-moving frames, respectively.
When the system of the points, moving with a constant proper acceleration $g$ is rigid [i.e., $dx^2+dy^2+dz^2=const$ in the co-moving rest frame measured in time coincidence $dt=0$], in the Lab-frame they will represent a family of hyperbolic trajectories with a constant interval $ds^2 =dX^\mu dX_\mu = dt^2(1+gz)^2-(dx^2+dy^2+dz^2)=const$. Thus, the family of the hyperbolic trajectories given by Rindler coordinates represents rigid dynamics.

\textit{Quantum dynamics of an accelerating electron.} As previously mentioned, the nonspreading wavepacket is closely related to the confined Dirac solution of Greiner  \cite{Greiner_1985}  for the electron in a constant gravitational field and, due to the equivalence principle, can be deduced from it.
In the chiral representation \cite{hestenes1975observables}, the eigenspinor of the Greiner's solution for a spin-up electron  reads:
\begin{align}\label{RindlerSpinor}
\psi_R=\frac{2\sqrt{2}\mathcal{N}e^{\pi\Omega/2}}{i\pi}e^{i\gamma^5\pi/4}\left(\begin{array}{c}
K_{i\Omega+1/2}(m u)
   \\
0
   \\
K_{i\Omega-1/2}(m u)
   \\
0
\end{array}
\right)e^{-i\Omega \eta},
\end{align}
where $K_\nu(x)$ is a Bessel function, $\mathcal{N}$ is a normalization constant, $m $ is the electron mass, $\gamma^5=i\gamma^0\gamma^1\gamma^2\gamma^3$, and $\Omega$ the eigenenergy.
 $(\eta,u)$ are defined as the comoving coordinates of an inertial observer momentarily at rest with respect to the electron. Hence, using the Rindler coordinates we have $\eta\equiv gt$ and $u\equiv z+1/g=\sqrt{(Z+1/g)^2-T^2}$. The spinor (\ref{RindlerSpinor}) can be cast in the following form, see Eq. (13) in Sec.~ \ref{B}:
\begin{align}\label{RindlerSpinor2}
\psi_R&=\frac{i\sqrt{2}\mathcal{N}}{\pi}e^{-\frac{\gamma^0\gamma^3}{2}w }\int_{-\infty}^\infty db \left(\begin{array}{c}
e^{-\frac{i\Omega+b}{2}}
   \\
0
   \\
e^{-\frac{i\Omega-b}{2}}
   \\
0
\end{array}
\right)
  e^{-im(T\cosh b-Z\sinh b)}
\end{align}
with the momentum parameterized by the rapidity $b$ as $p=m\sinh b$ and $w=\tanh^{-1}(T/Z)$.
The wavefunction of Eq.~(\ref{RindlerSpinor}) is an eigenstate  and is confined in the coordinate $u$. Note that only for gravitational fields can an accelerated electron be described as a superposition of plane waves. This is a direct consequence of the equivalence principle. In fact, only gravity induced acceleration can be transformed away by a coordinate transformation in the immediate vicinity of the particle.

The  confined solution for the eigenstate $\psi_R$  to the free Dirac equation  with respect to the accelerated frame $(\eta,u)$ of Eq.~(\ref{RindlerSpinor2})  can be mimicked by a superposition $\psi$ of the Dirac solutions for a free electron with respect to the Lab-frame $(T,Z)$ (see Sec.~\ref{B}):
\begin{eqnarray}\label{psiR}
\psi_R=e^{-\frac{\gamma^0\gamma^3}{2}\tanh^{-1}\left(\frac{T}{Z}\right)}\psi,
\end{eqnarray}
where the free wavepacket $\psi$ should have a momentum chirp via the phase $\varphi(b)=-\alpha b$, with $\alpha=\Omega$, according to Eq.~(\ref{RindlerSpinor2}). When additionally we use the momentum distribution in the free wavepacket  $h(p)=e^{-aE_p}$, with the constant $a>0$ characterizing the momentum spread of the wavepacket, we get the following  dispersionless free spinorial wavepacket (hereinafter, overbar stands for the correspondingly dimensionless parameters): 
\begin{align}\label{superp0}
\psi (T,X)&=\mathcal{N}\left(
\begin{array}{c}
F_{i\alpha-1/2}(\bar{\zeta})
   \\
0
   \\
F_{i\alpha+1/2}(\bar{\zeta})
   \\
0
\end{array}
\right),
 \end{align}
where $\bar{\zeta}=i\sqrt{(\bar{a}+i\bar{T})^2+\bar{Z}^2}$ and $F_{i\alpha\pm1/2}(\bar{\zeta})=2\left(\frac{i\bar{a}-\bar{T}-\bar{Z}}{i\bar{a}-\bar{T}+\bar{Z}}\right)^{\pm1/4+i\alpha/2} K_{\pm1/2+i\alpha}(\bar{\zeta})$. While the wavepacket (\ref{superp0}) is discussed already in Ref.~\cite{Kaminer_2015}, the emphasis here is its direct relation to the nonspreading concept.

\textit{Nonspreading wave packet in a strong laser field.} Our aim is to construct a solution of the Dirac equation in a laser field in the form of a wavepacket and to show, using the CRDI technique, that it represents a nonspreading spinor in the local rest frame of the electron.
We construct the desired wavepacket from the Volkov solutions $\psi_p(T,X)$ for an electron in a plane wave laser field $eA^\mu=(0,\dot{f}_1(\xi),\dot{f}_2(\xi),0)$: 
\begin{align}\label{superp}
\psi_L(T,X)&=\frac{1}{(2\pi )^{1/2}}\int_{-\infty}^\infty \frac{dp}{2E_p} f(p)\psi_p(T,X)
 \end{align}
where 
\begin{align}\label{PsiM}
\psi_p(T,X)=\left(\mathbf{1}+\frac{n\!\!\!/\wedge \mathbf{A}\!\!\!/}{n^\mu p_\mu}\right)\mathbf{u}e^{-i(E_pT-pZ-\Phi)},
\end{align}
with $n^\mu =(1,0,0,1)$, $p_\mu=(E_p,0,0,-p)$, $E_p =\sqrt{m^2+p^2}$, $\mathbf{A}\!\!\!/=\gamma^\mu eA_\mu$, $n\!\!\!/\wedge \mathbf{A}\!\!\!/=(n\!\!\!/ \mathbf{A}\!\!\!/-\mathbf{A}\!\!\!/n\!\!\!/)/2$, $\mathbf{u}$ is the leftmost column of the boost matrix $\mathcal{B}=\sqrt{\frac{E_p+m}{2E_p}}\left(\mathbf{1}+\frac{\gamma_0\gamma^3p}{E_p+m}\right)$, $\Phi=-\frac{1}{2\omega(E_p-p)}\int_0^\xi[\dot{f}_1(\phi)^2+\dot{f}_2(\phi)^2]d\phi$, and the laser field phase $\xi=\omega(T-Z)$. For the superposition coefficients in Eq.~(\ref{superp}) we use those which yield the free wavepacket $\psi$, see Eqs.~(\ref{RindlerSpinor2})-(\ref{psiR}). After performing the  change of variables $p=m\sinh b$
\begin{align}\label{RindlerSpinorLaser}
\psi_L&=\int_{-\infty}^\infty db\mathcal{N}f(b) \left(\begin{array}{c}
e^{-b/2}
   \\
e^{\frac{b}{2}}[\dot{f_1}(\xi)+i\dot{f_2}(\xi)]/m
   \\
e^{b/2}
   \\
0
\end{array}
\right)\nonumber\\
 &\times e^{-im(T\cosh b-Z\sinh b)-i e^b\Phi}
\end{align}
the closed expression for the integral in (\ref{RindlerSpinorLaser}) is
\begin{align}\label{superp2}
\psi_L(T,X)&=\mathcal{N}\left(
\begin{array}{c}
F_{i\alpha-1/2}(\bar{\zeta}')
   \\
F_{i\alpha+1/2}(\bar{\zeta}')[\dot{f_1}(\xi)+i\dot{f_2}(\xi)]/m
   \\
F_{i\alpha+1/2}(\bar{\zeta}')
   \\
0
\end{array}
\right),
 \end{align}
where $\bar{\zeta}'=i\sqrt{(\bar{a}+i\bar{T}')^2+\bar{Z}'^2}$, $Z'=Z-\Phi,\quad T'=T+\Phi$ and $F_{i\alpha\pm1/2}(\bar{\zeta}')=2\left(\frac{i\bar{a}-\bar{T}'-\bar{Z}'}{i\bar{a}-\bar{T}'+\bar{Z}'}\right)^{\pm1/4+i\alpha/2} K_{\pm1/2+i\alpha}(\bar{\zeta}')$.  

Let us  transform  the spinorial wavepacket (\ref{superp2}) to the rest frame, which is defined as the space-time dependent frame in which the spatial components of the electron's four-current vanish at the given space-time point, and demonstrate its nonspreading property. In the free-electron case, such Lorentz transformation is the matrix $e^{-\gamma^0\gamma^3w/2}$ on the left of the spinor (\ref{RindlerSpinor2}). An equivalent transformation is now needed for the case in which the electron is interacting with a plane wave field.
In order to construct the desired Lorentz transformation, we make use of the CRDI technique. In the Hestenes formulation (see, for instance, section 3 of  Ref. \cite{hestenes1974proper}), the spinor (\ref{PsiM}) can be written as
\begin{align}\label{PsiM2}
\Psi=e^{\frac{n\!\!\!/\wedge \mathbf{A}\!\!\!/}{n^\mu p_\mu}}\mathcal{B}e^{-\gamma_2\gamma_1(E_pT-pZ-\Phi)}.
\end{align}
As  discussed in \cite{Campos_2022} the matrix $e^{\frac{n\!\!\!/\wedge \mathbf{A}\!\!\!/}{n^\mu p_\mu}}\equiv \mathcal{R}$ is, in fact, a Lorentz transformation.  In the chiral representation it is given by
 \begin{align}\label{NullRpre}
\mathcal{R}=\begin{pmatrix}
    1&0&0 & 0\\
    \frac{(\dot{f}_1(\xi)+i\dot{f}_2(\xi))}{E_p-p}&1&0&0\\
    0&0&1&-\frac{(\dot{f}_1(\xi)-i\dot{f}_2(\xi))}{E_p-p}\\
  0   &0&0& 1
    \end{pmatrix}.
\end{align}
However, the Lorentz transformation (\ref{NullRpre}) is valid only for the wavefunction (\ref{PsiM2}) but not (\ref{superp2}). Moreover, it does not account for the transformation to the Rindler (accelerated) frame. In order to encompass both transformations, we start with the following ansatz 
 \begin{align}\label{NullR4}
\bar{\mathcal{R}}=e^{-\frac{\gamma^0\gamma^3\eta'}{2}}\begin{pmatrix}
    1&0&0 & 0\\
    -d^\ast\omega(\dot{f}_1(\xi)+i\dot{f}_2(\xi))&1&0&0\\
    0&0&1&d\omega(\dot{f}_1(\xi)-i\dot{f}_2(\xi))\\
  0   &0&0& 1
    \end{pmatrix}
\end{align}
where $d^\ast$ and $\eta'$ are free functions to be found by the requirements that the resulting spinor is of the same form as Eq.~(\ref{RindlerSpinor2}) and that the electron's current vanishes. These requirements are fulfilled by the following functions
\begin{align*}
d^\ast=\sqrt{\frac{a+i \left(c T'+Z'\right)}{a+i \left(c T'-Z'\right)}}\frac{
   K_{i \alpha -\frac{1}{2}}\left(\kappa \bar{\zeta}'\right)}{2 m \omega  K_{i
   \alpha +\frac{1}{2}}\left(\kappa \bar{\zeta}'\right)}
\end{align*}
and
 \begin{align*}
 \eta'&=\frac{1}{2}\mbox{ln}\left(q\frac{K_{1/2-i\alpha}(\kappa \bar{\zeta}'^\ast)K_{1/2+i\alpha}(\kappa \bar{\zeta}')}{K_{1/2-i\alpha}(\kappa \bar{\zeta}')K_{1/2+i\alpha}(\kappa \bar{\zeta}'^\ast)}\right),\\
 q&=\sqrt{\frac{a^2+(cT'+Z')^2}{a^2+(-cT'+Z')^2}}
 \end{align*}
with superscript $^\ast$ standing for complex conjugation.
Applying $\bar{\mathcal{R}}$ to (\ref{superp}), $\bar{\mathcal{R}}\,\psi_L$, leads to the following spinor describing the electron in its rest frame:
 \begin{align}\label{ComovingM}
 &\bar{\mathcal{R}}\,\psi_L\equiv\bar{\psi}_R=\mathcal{N}e^{-\frac{\gamma^0\gamma^3}{2}\eta'}\left(
\begin{array}{c}
F_{i\alpha-1/2}(\bar{\zeta}')
   \\
0
   \\
F_{i\alpha+1/2}(\bar{\zeta}')
   \\
0
\end{array}
\right).
 \end{align}
The final step of the transformation consists of the following coordinate transformation $ Z'=Z-\Phi,\quad T'=T+\Phi$.

One now needs to prove that the constructed wave function $\bar{\psi}_R$  obeys the Dirac equation in the plane wave field given by $\mathbf{A}\!\!\!/$. In order to do so one must do the following:
First construct the vierbein $e^\alpha_\mu=\frac{1}{4}\Tr[\bar{\mathcal{R}}^{-1}\gamma^\alpha\bar{\mathcal{R}}\gamma_\mu]$, $e^\mu_\alpha=\frac{1}{4}\Tr[\bar{\mathcal{R}}^{-1}\gamma^\mu\bar{\mathcal{R}}\gamma_\alpha]$, which transform the $\gamma$-matrices in the lab frame to the new $\gamma$-matrices
 $ \tilde{\gamma}^\alpha=e^\alpha_\mu\gamma^\mu,\quad  \tilde{\gamma}_\alpha=e_\alpha^\mu\gamma_\mu $. After the transformation the wavepacket $\bar{\psi}_R$ satisfies the following Dirac equation
 $$
 i\tilde{\gamma}^\mu\nabla_\mu\bar{\psi}_R-\tilde{\gamma}_\mu eA^\mu\bar{\psi}_R-m\bar{\psi}_R=0,
 $$
where $\nabla_\mu=\partial/\partial X^\mu+\mathbf{\Omega}_\mu$, $X^\mu=(T,X,Y,Z)$, and describes the electron in its rest frame. The matrix $\mathbf{\Omega}_\mu$ is the spinor connection and is given by
$2\mathbf{\Omega}_\mu=\Omega_{ij\mu}\boldsymbol{\sigma}^{ij}$, $2\boldsymbol{\sigma}_{ij}=\gamma_i\gamma_j$ and $\Omega^i_{j\mu}=-e^\nu_je^i_\sigma e^\sigma_a\partial_\mu e^a_\nu$.
The wavepacket in the rest frame will be nonspreading when the spinor of Eq.~(\ref{superp}) is represented in exactly the same form as the free spinor of Eq.~(\ref{kaminerandlaser4}). This is achieved by the consecutive application of 
the coordinate transformation $ Z'=Z-\Phi,\quad T'=T+\Phi$. With the change of vierbein 
  $
 e'^\alpha_\mu=\frac{\partial X'^\alpha}{\partial X^\mu},\, e'^\mu_\alpha=\frac{\partial X^\mu}{\partial X'^\alpha}
 $, and with the new gamma-matrices  $ \gamma'^\alpha=e'^\alpha_\mu\tilde{\gamma}^\mu,\quad  \gamma'_\alpha=e'^\mu_\alpha\tilde{\gamma}_\mu$,
the transformed spinor (\ref{ComovingM}) satisfies the Dirac equation in the new frame
  $$
 i\gamma'^\mu\nabla'_\mu\bar{\psi}_R-\tilde{\gamma}_\mu eA^\mu(\xi')\bar{\psi}_R-m\bar{\psi}_R=0
 $$
with $\nabla'_\mu=\partial/\partial X'^\mu+\mathbf{\Omega}_\mu$. Note also that $ eA^\mu(\xi')=eA^\mu(\xi(\xi'))$ since $\xi'=\xi+2\omega \Phi(\xi)$.
The variable $\xi(\xi')$ is then given by inverting the coordinate transformation. 

 \begin{figure}[b]
 \includegraphics[width=0.23\textwidth]{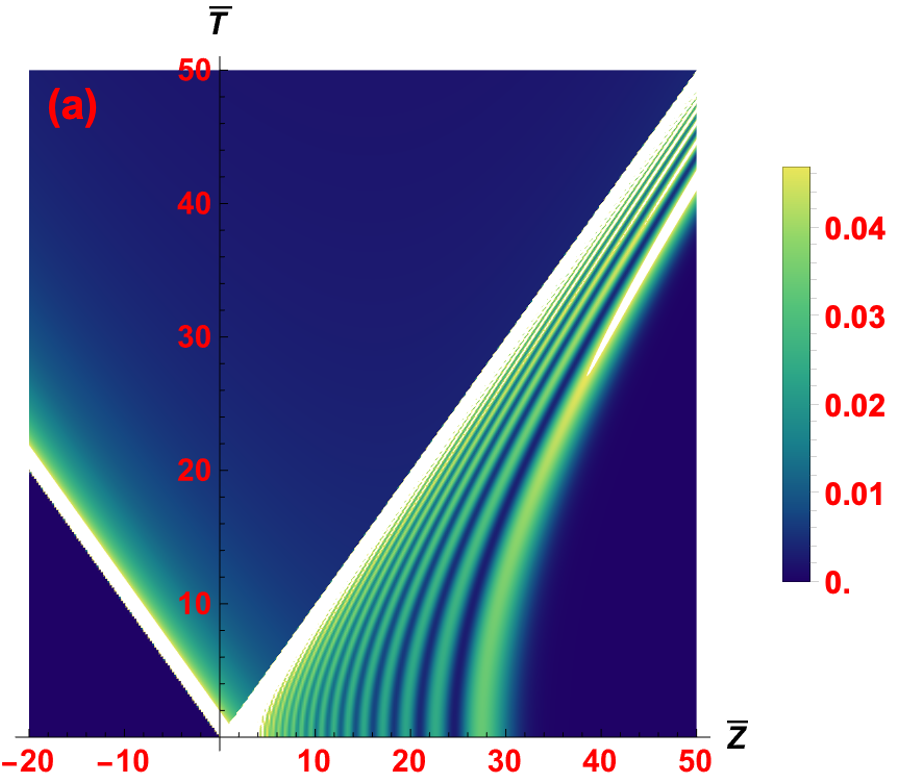}
  \includegraphics[width=0.23\textwidth]{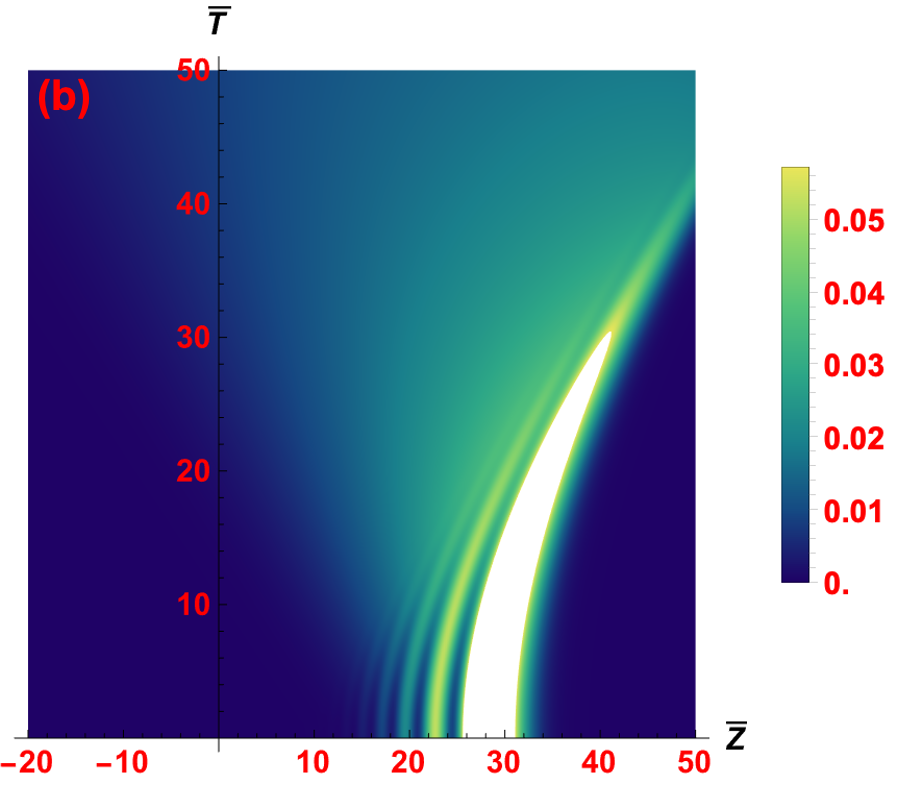}
  \includegraphics[width=0.23\textwidth]{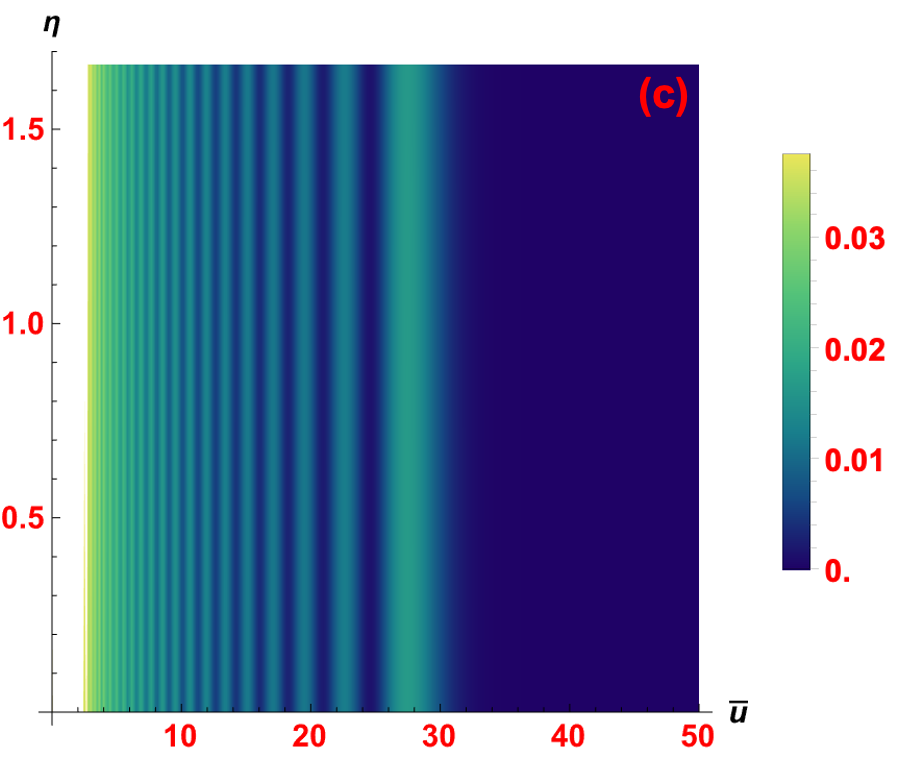}
  \includegraphics[width=0.23\textwidth]{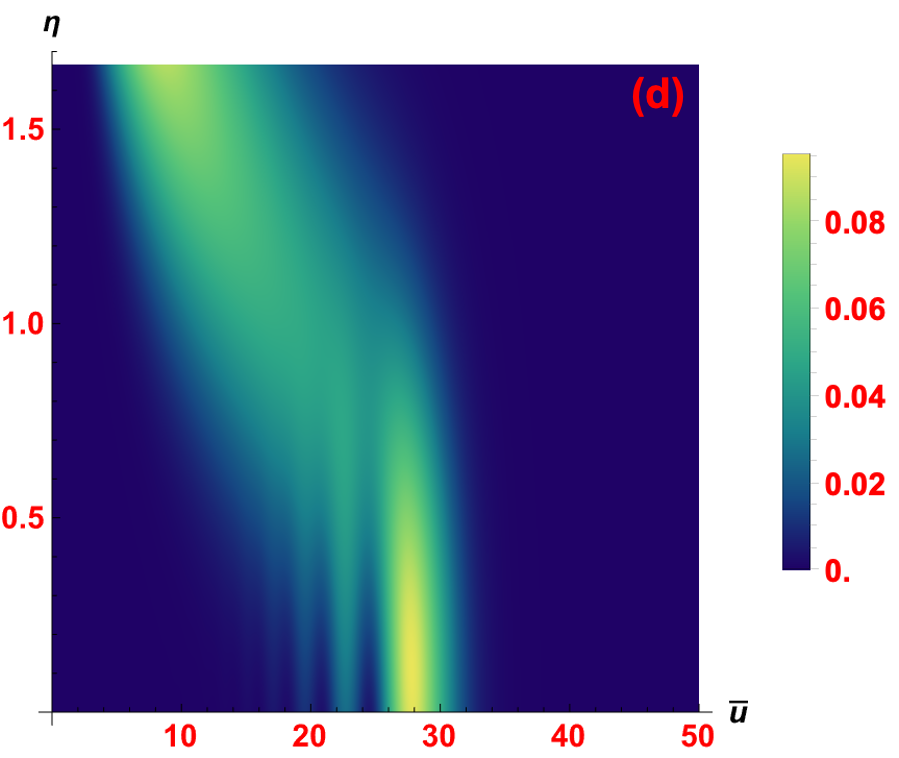}
   \caption{Spacetime profile of the electron density for the free electron modulated wavepacket of Eq.~(\ref{kaminerandlaser4}): (a,b) in the Lab-frame; (c,d)  in the accelerated frame (i.e., in Rindler coordinates); (a,c) $\alpha=30$ and $\bar{a}=0.005$; (b,d) $\alpha=30$ and $\bar{a}=2$. Note that the Rindler coordinates only cover the region outside the light-cone to the right.}
      \label{Densities}
\end{figure}

Thus, the constructed wavepacket of the electron in a laser field in the form of Eq.~(\ref{superp}) [or Eq.~(\ref{superp2})] coincides, up to a boost, see Eq.~(\ref{ComovingM}), with the free self-accelerating nonspreading wavepacket [cf. Eq.~(\ref{superp0})] in the local rest frame of the electron at each $(Z',T')$.
Note that the exact Lorentz transformation of the electron Dirac wave function in a laser field to the electron rest frame  is essentially facilitated by application of the CRDI technique \cite{Campos_2017,Campos_2022}.

There is an important deviation of the nonspreading wave packet Eq.~(\ref{ComovingM}) from the accelerating electron solution Eq.~(\ref{RindlerSpinor}). While in the latter $a=0$, the former has a finite size of the wavepacket $a\neq 0$ which has essential implications. To discuss this, consider the spinor (\ref{ComovingM}) in the lab frame (hereinafter we drop the primes in order to simplify the notation)
\begin{align}
\label{kaminerandlaser4}
\psi&=e^{\frac{\gamma^0\gamma^3}{2}\eta}\bar{\psi}_R=\mathcal{N}\left(\begin{array}{c}F_{i\alpha+1/2}(\bar{\zeta}) \\0   \\F_{i\alpha-1/2}(\bar{\zeta})   \\0\end{array}\right),
\end{align}
 The impact of the wave packet size $a$  is given by the  prefactor in $F_{i\alpha\pm1/2}(\bar{\zeta})$, which we analyze next.

The spacetime profile of the electron wavepacket is presented in Fig. \ref{Densities}. The panels (a,b) show the distribution in the Lab-frame coordinates $(T,Z)$, while (c,d) in the accelerated rest frame with Rindler coordinates ($\eta, u$). In the Lab-frame, the wavepacket is separated into two parts: inside the light-cone with normal spreading, and outside the light-cone  
(defined as Region~I in Ref.~\cite{Greiner_1985}) 
representing the nonspreading wavepacket. The latter shows interference fringes, each lobe  corresponding to a hyperbolic trajectory. Such a feature is entirely due to the chirping parameter $\alpha$. From the rest frame perspective [Fig. \ref{Densities}(c,d)] the wavepacket is nonspreading (the width of the wavepacket at each instant of the Rindler's time $\eta$  remains constant), while in the Lab-frame the width of the wavepacket is contracting with time $T$ according to the Lorentz-transformation.

\begin{figure}[t]
   \includegraphics[width=0.23\textwidth]{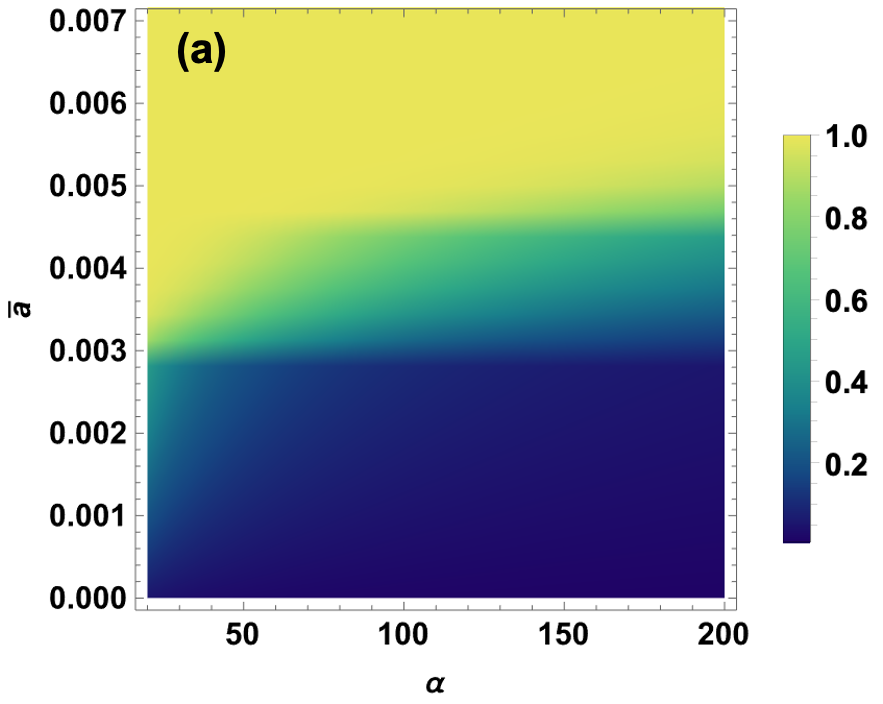}
   \includegraphics[width=0.23\textwidth]{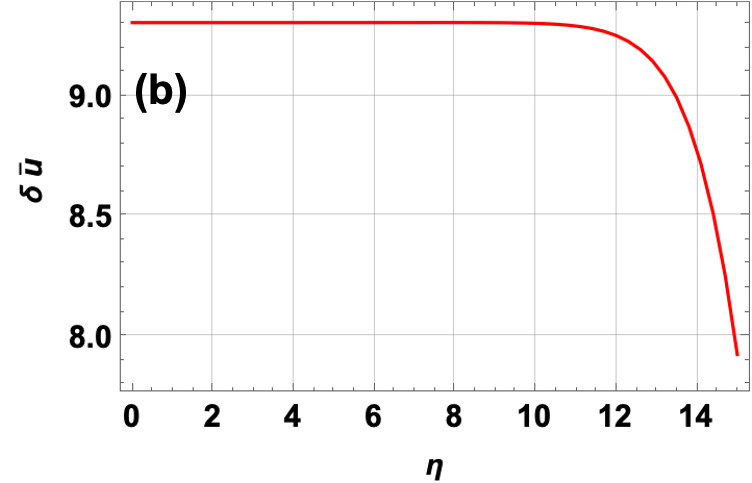}
  \caption{ (a) The asymmetry ratio ${\cal A}=(|\bar{\psi}_R|^2-|\psi|^2|)/(|\bar{\psi}_R|^2+|\psi|^2|)$ vs the wavepacket chirping parameter $\alpha$ and the size parameter $\bar{a}$ for $T=1$ fs; (b)  The wavepacket size $\delta\bar{u}$ vs $\eta$ for $\alpha=40$ and $\bar{a}=10^{-6}$. }
       \label{asymmetry}
\end{figure}

There is a significant effect stemming from the value of the wavepacket size parameter $a$, cf.  Panels (a,c) with (b,d) in Fig.~\ref{Densities}. During evolution, the  nonspreading part of the wavepacket is gradually leaking out into the normal one. The parameter $a$ controls the balance between the nonspreading and normal parts of the wavepacket, see section \ref{C}, and determines the lifetime of the  nonspreading part of the wavepacket. Such leaking is responsible for washing out the interference fringes: the smaller the value of $a$, the slower is the interference fringes extinction [Fig. \ref{Densities}]. There is no extinction in the case of the accelerating electron solution of Eq.~(\ref{RindlerSpinor}) with $a=0$. We can estimate the lifetime of the nonspreading wave packet using asymptotic expressions of the wavefunctions at $\eta \gg 1$, see section \ref{C}: 
 $|\psi|^2\approx\ \exp[-2 \left(e^{\eta}  \bar{a} \bar{u}\right)^{1/2}-\frac{\pi \alpha }{2}]/\sqrt{2}\bar{u}$,  $|\psi_R|^2\approx \exp[-2\bar{u}+\frac{\pi \Omega }{2}]/\sqrt{2} \bar{u}$.
 Both asymptotic expansions will coincide, if  $ e^{\eta} \bar{a} \sim \bar{u}$, or $\bar{Z}-\bar{T}\gtrsim \bar{a}$. Taking into account that the equation for the rightmost hyperbolic trajectory ($\bar{Z}_0\approx\alpha$) as a function of time is $\bar{Z}(\bar{T})\approx\sqrt{\alpha^2+\bar{T}^2}$, we have an estimate for the lifetime of the nonspreading wavepacket:
\begin{align}\label{avalue}
T_l\lesssim (\alpha^2-\bar{a}^2)/(2\bar{a}m),
\end{align}
which indicates that large $\alpha$ and small $\bar{a}$ are beneficial for the extension of the lifetime. For instance, $T_l\lesssim 1$ fs when using $\alpha=30$ and $\bar{a}=0.001$.

We also analyze the balance of the nonspreading and normal parts of the wavepacket introducing the asymmetry parameter via ${\cal A}=(|\bar{\psi}_R|^2-|\psi|^2|)/(|\bar{\psi}_R|^2+|\psi|^2|)$, calculating densities of these  parts via Eqs.~(\ref{RindlerSpinor2}), (\ref{kaminerandlaser4}), see section \ref{D} for a definition. The example of the dependence of ${\cal A}$ on the parameters $\alpha$ and $\bar{a}$ is shown in Fig.~\ref{asymmetry}(a) for $T=1$ fs. The wavepacket is nonspreading if $|\psi|^2\approx |\bar{\psi}_R|^2|$, i.e. at ${\cal A}\rightarrow 0$, while at ${\cal A}\rightarrow 1$,  the nonspreading wave packet is fully extinguished $|\psi|^2\rightarrow 0$.  ${\cal A}$ is very sensitive to $\bar{a}$. Smaller $\bar{a}$ is preferred for ${\cal A}\rightarrow 0$, however, the larger $\alpha$ allows larger $\bar{a}$ at a given ${\cal A}$ [Fig.~\ref{asymmetry}(a)].

We numerically evaluated the wavepacket spatial size via the accelerated frame spinor, see Sec.~\ref{D}, i.e., considering only the parts outside the light-cone. The standard deviation $ \delta\bar{u}=\sqrt{\langle\bar{u}^2\rangle-\langle\bar{u}\rangle^2}$ is calculated with $\bar{a}=10^{-6}$ and $\alpha=40$ [Fig.~\ref{asymmetry}(b)]. With the rightmost hyperbolic trajectory the transformation between the time in the accelerated frame and in the Lab-frame is then $\bar{T}=\alpha\sinh\eta$. As seen in Fig.~\ref{asymmetry}(b), the wavepacket spreading $ \delta\bar{u}$ stays constant up to $\eta\approx 11$ which corresponds to the  Lab-time $T\approx 2$ fs.

Considering that the fringes of the self-accelerating part of the wavepacket must last for at least one period of the laser field, let us estimate the maximum value for the parameter $\bar{a}$. For a full cycle of the laser field in the electron's rest frame, one has $\omega(T-Z)=2\pi$. The latter combined with the condition $|\bar{Z}-\bar{T}|\gtrsim\bar{a}$, we have $a \lesssim  \lambda'$, where $\lambda'$ is the laser wavelength in the electron rest frame.

We consider  applications of the nonspreading  relativistic wavepackets to a laser-driven  collider  \cite{Meuren_2015}. Here electrons and positrons are created from vacuum by high-energy gamma-photons counterpropagating an ultrastrong laser field. They are accelerated by the laser field and collide within a cycle of the  field.  The rest frame of the created pairs depends on the $\gamma$-photon energy ($\Omega_0$) and the laser strong field parameter ($a_0\equiv eE_0/(m\omega)$, with the laser field amplitude $E_0$). We estimated \ref{F} that at $\Omega_0\sim 1$ GeV and $a_0=10^2$ (the laser intensity of $10^{22}$ W/cm$^2$), the rest frame of the pair moves with $\gamma\approx 30$. The laser period in this frame $T'=T_L/\gamma\sim 3\times 10^{-2}$~fs (with the laser period $T_L$ in the Lab-frame) which is less than the wavepacket leaking time $T_l\sim 1$~fs (for $\alpha=30$ and $\bar{a}=0.001$), i.e., the recollision time is short enough to maintain the nonspreading character of the wavepacket. The next point is how to create the nonspreading wavepacket (see Sec.~ \ref{F}). For the latter the specially tailored momentum chirping of the wavepacket given by the phase $\varphi(p)$ is essential. This chirping induces a spatial shift of each momentum component in the laser field $\delta x (p)=\partial \varphi(p)/\partial p$. The created wavepacket of the electron (positron) will be chirped if the particle with the corresponding momentum value is created with the corresponding spatial delay $\delta x (p)$. The created particle  momentum in the Lab-frame is determined either by the laser field intensity, or by the $\gamma$-photon energy. Tailoring specifically the laser intensity in space according to the function $\delta x (p)$, one can achieve chirping of the created wavepackets of the electron and positron. Another possibility is to use a chirped $\gamma$-photon beam. The approach based on the Dirac equation is applicable here because radiation reaction is negligible, as shown in Sec.~\ref{F}, for the typical parameters of the laser-driven collider.

\textit{Conclusion.} We have shown the existence of nonspreading relativistic wavepackets in a laser field, which in the  local rest frame of the electron  is similar to  a self-accelerating nonspreading free wavepacket.
We have established that there is a finite lifetime for the self-accelerating  wavepacket and found that the wavepacket chirping and extension parameters impose strict restrictions on the lifetime duration. The nonspreading feature of the relativistic wave packet represents the essential property to permit an efficient laser-driven high-energy collider. 

Moreover, the nonspreading free electron wavepacket is obtained via mimicking the confined Dirac eigenstate in a constant gravitational field. This idea based on the equivalence principle can be further developed, constructing different free electron wavepackets emulating  electron bound states in a gravitational field of various configurations. This would provide a new avenue in laboratory astrophysics/cosmology: to investigate particle quantum dynamics in gravitational fields via the dynamics of their physical counterpart with specially engineered free electron wavepackets \cite{Wollenhaupt_2002,Feist_2015,Kealhofer_2016,Shiloh_2022,Dienstbier_2023}. 
 
 
\appendix
\begin{widetext}

\section{Construction of a family of rigid relativistic coordinate systems}\label{A}
In what follows, greek indices run from $0$ to $3$ while latin indices run from $1$ to $3$. 
In line with the concept of Born rigidity, here we show how to construct a rigid reference system. Consider a particle in arbitrary motion relative to an inertial system $I$; the particle's coordinates with respect to $I$ are $X^\alpha=(cT,X,Y,Z)$ where $c$ is the speed of light in vacuum. The particle's time track may be described by the equations $X^\alpha=f^\alpha(\tau)$, 
$\tau$ being the proper time of the particle. Consider now another reference system $R$ attached to the particle which is uniformly accelerated with respect to $I$; the axis of $R$ should always be parallel to that of $I$, the particle being always situated at its origin. Let the coordinates following the particle in its motion relative to $R$ be $x^\alpha=(ct,x,y,z)$. At any moment there exists an inertial coordinate system $I'$, momentarily at rest with respect to the particle, whose coordinate axes coincide with those of $R$. Hence, we have $x'_i=x_i$, $x'_0=0$ and $\tau=t$. The transformation connecting the coordinates $X^\alpha$ with $x^\alpha$ is
\begin{align}\label{System1}
X^\alpha=f^\alpha(t)+x^ie_i^{\,\alpha}(t),
\end{align}
where $e_\nu^{\,\alpha}(t)$ is an orthonormal frame for an accelerated observer that obeys the following equation
$$
 \frac{de_\mu^\nu}{dct}=\Omega^\nu_{\,\,\beta}e^\beta_\mu,
$$
where $\Omega^\nu_{\,\,\beta}=u^\nu\dot{u}_\beta-u_\nu\dot{u}^\beta$ and $u_\mu=\dot{f}_\mu(t)=df_\mu(t)/dct$. Differentiation of $X^\alpha$ gives $dX^\alpha=(u^\alpha+x^i\dot{e}_i^{\,\alpha}(t))cdt+dx^ie_i^{\,\alpha}(t)$. From the following properties $u^\alpha u_\alpha=1$, $\dot{u}_\beta \dot{u}^\beta=-g_ig^i$ and $\dot{u}_\beta u^\beta=0$ we get
\begin{align}\label{accSystem}
 ds^2=c^2dt^2(1+g_ix^i/c^2)^2-(dx^2+dy^2+dz^2),
 \end{align}
 where $g_i(t)=e_i^{\,\alpha}\dot{u}_\alpha$ are functions of $t$ only, being completely determined by the motion of the origin of the system of coordinates $x^\alpha$ relative to the system $X^\alpha$. Considering the line element (\ref{accSystem}) the corresponding system of reference is rigid since the distance between two reference points $(x,y,z)$ and $(x+dx,y+dy,z+dz)$ is given by $d\sigma^2=dx^2+dy^2+dz^2$. In fact, the space geometry is even Euclidean; thus $(x,y,z)$ are cartesian space coordinates. 
 
 Now consider that the origin $O$ of the system $x^\alpha$ is moving in the $Z$-axis direction of the $X^\alpha$ system. From Eqs. (\ref{System1}) we have
 \begin{align}\label{Zmotion}
 &X=x,\, Y=y,\, Z=c\int_0^t\sinh(\theta(t))dt+z\cosh(\theta(t)),\nonumber\\
&T=\int_0^t\cosh(\theta(t))dt +\frac{z}{c}\sinh(\theta(t)).
 \end{align}
 For the vector $g_i$ we get $\mathbf{g}=(0,0,g(t))$, $g(t)=cd\theta/dt$. Hence, Eq. (\ref{accSystem}) becomes
 \begin{align}\label{accSystem2}
  ds^2=c^2dt^2(1+gz/c^2)^2-(dx^2+dy^2+dz^2).
 \end{align}
 In particular, if the motion of the origin $O$ is hyperbolic, then $\theta(t)=gt/c$ thus making $g$ a constant. The transformation equations then reduce to
  \begin{align}\label{Zmotion2}
 &X=x,\, Y=y,\, Z=\frac{c^2}{g}\left(\cosh(gt/c)-1\right)+z\cosh(gt/c),\nonumber\\
&T=\frac{c}{g}\sinh(gt/c) +\frac{z}{c}\sinh(gt/c).
 \end{align}
 Let us examine if the reference system $R$ corresponding to the coordinates $x^\alpha$ will appear as rigid with respect to the observer $A$ in the inertial frame $I$.
 By elimination of the variable $t$ from Eqs. (\ref{Zmotion2}) we obtain
 $$
 Z=\frac{c^2}{g}\left(\sqrt{(1+gz/c^2)^2+g^2T^2/c^2}-1\right),\, X=x,\, Y=y.
 $$
 The velocity of the reference points relative to $I$ at time $T$ is thus
 \begin{align}\label{V}
 \frac{v}{c}=\frac{dZ}{cdT}=\frac{gT}{c\sqrt{(1+gz/c^2)^2+g^2T^2/c^2}}=\tanh(gt/c).
 \end{align}
 Since the velocity $v$ of the frame $R$ from the point of view of $A$ also depends on $z$, $R$ will not appear as rigid with respect to $I$. In fact, the difference between two reference points $(x,y,z)$ and $(x,y,z+dz)$
 measured by $A$ is found to be
 \begin{align*}
 dZ&=\frac{(1+gz/c^2)dz}{\sqrt{(1+gz/c^2)^2+g^2T^2/c^2}}=\frac{dz}{\cosh(gt/c)}\\
 &=\sqrt{1-v^2/c^2}dz.
 \end{align*}
 Hence, from $A$'s perspective each part of the accelerated frame $R$ undergo a Lorentz contraction.
 
Going back to the motion of the particle in the accelerated frame $R$ from the point of view of $A$, consider the velocity $v$ for the particle located permanently at position $z$, that is, it is at rest with respect to $R$.
By definition, the proper time of the particle can be calculated from Eq. (\ref{accSystem2}) as 
\begin{align}\label{propertime}
cd\tau&=\sqrt{c^2dt^2(1+gz/c^2)^2-dx^2-dy^2-dz^2}\\
&=cdt(1+gz/c^2),\\
\tau&=(1+gz/c^2)\int_0^t dt=(1+gz/c^2)t,
\end{align}
given that $dx/dt=dy/dt=dz/dt=0$.

\comm{Going back to the motion of the particle in the accelerated frame $R$ from the point of view of $A$, consider the velocity $v$ for the particle located permanently at position $z$, that is, it is at rest with respect to $R$.
By definition, the proper time of the particle can be calculated from Eq. (\ref{accSystem2}) as 
\begin{align}\label{propertime}
cd\tau&=\sqrt{c^2dt^2(1+gz/c^2)^2-dx^2-dy^2-dz^2}\\
&=cdt(1+gz/c^2),\\
\tau&=(1+gz/c^2)\int_0^t dt=(1+gz/c^2)t,
\end{align}
given that $dx/dt=dy/dt=dz/dt=0$. }
\section{Construction of the free Dirac spinor}\label{B}
\subsection{Exact solution of the Dirac equation for an electron in a frame moving with constant four-acceleration.}
Here we rewrite the spinor solution for an electron in a homogeneous and constant gravitational field as an integral in order to demonstrated its relation with the self-acceleration spinor.
The concept described here holds for any spatial dimension. Let us first describe the case of a single spatial dimension as discussed in the previous section.
In the Chiral representation, the Dirac spinor for a spin-up electron in a reference frame undergoing constant four-acceleration is
\begin{align}\label{PreRindlerSpinor}
\psi_R=\sqrt{2}\mathcal{N}\left(\begin{array}{c}
H_{i\Omega+1/2}^{(1)}(imcu/\hbar)
   \\
0
   \\
H_{i\Omega-1/2}^{(1)}(imcu/\hbar)
   \\
0
\end{array}
\right)e^{-i\Omega \eta},
\end{align}
which by using the connection formula  $H_{\nu}^{(1)}(ix)=\frac{2K_\nu(x)}{\pi ie^{i\pi\nu/2}}$ and defining $\kappa=mc/\hbar$ can be rewritten as
\begin{align}\label{RindlerSpinorB}
\psi_R=\frac{2\sqrt{2}\mathcal{N}e^{\pi\Omega/2}}{i\pi}e^{i\gamma_5\pi/4}\left(\begin{array}{c}
K_{i\Omega+1/2}(\kappa u)
   \\
0
   \\
K_{i\Omega-1/2}(\kappa u)
   \\
0
\end{array}
\right)e^{-i\Omega \eta}
\end{align}
where $K_\nu(x)$ is a Bessel function, $\mathcal{N}$ a normalization constant, $\Omega>0$ the electron's kinetic energy, $\gamma_5=\gamma^5=i\gamma^0\gamma^1\gamma^2\gamma^3$ and $g$ the constant four-acceleration proper length. The $(\eta,u)$ are defined as the comoving coordinates of an inertial observer momentarily at rest with respect to the electron. Hence, from Eqs. (\ref{Zmotion2}) after shifting the origin we have $\eta=gt/c$, $u=z+c^2/g$, $Z=u\cosh \eta$ and $cT=u\sinh \eta$. Let us massage Eq. (\ref{RindlerSpinorB}) a bit more to gain intuition on how to build it from a wavepacket. First, note that
$$
K_\nu(x)=\frac{1}{2}\int_{-\infty}^\infty e^{-x\cosh t+\nu t}dt,
$$
along with $\cosh(t)=i\sinh(t-i\pi/2)$. Combining both identities, (\ref{RindlerSpinorB}) becomes
\begin{align}\label{RindlerSpinor2B}
\psi_R=\frac{\sqrt{2}\mathcal{N}}{i\pi}\int_{-\infty}^\infty dt e^{-i\kappa u\sinh t}\left(\begin{array}{c}
e^{(i\Omega+1/2)t}
   \\
0
   \\
e^{(i\Omega-1/2)t}
   \\
0
\end{array}
\right)e^{-i\Omega \eta}
\end{align}
Since $Z=u\cosh \eta$, $cT=u\sinh \eta$, by defining the momentum $p=mc\sinh b$ for $b$ real and making the change of coordinates $t=\eta-b$ in (\ref{RindlerSpinor2B}), we finally have
\begin{align}\label{RindlerSpinor2aA}
\psi_R&=\frac{i\sqrt{2}\mathcal{N}}{\pi}e^{-\frac{\gamma^0\gamma^3}{2} \tanh^{-1}\left(\frac{cT}{Z}\right)}\int_{-\infty}^\infty \left(\begin{array}{c}
e^{-b/2}
   \\
0
   \\
e^{b/2}
   \\
0
\end{array}
\right)\nonumber\\
&\times e^{-i\kappa(cT\cosh b-Z\sinh b)-i\Omega b}db.
\end{align}
The spinor (\ref{RindlerSpinor2aA}) is the desired result.
\subsection{The Dirac equation in the accelerated frame}
Here we show how the spinor discussed in the previous section is connected with the self-accelerated spinor in the lab frame by a Lorentz transformation.
From the coordinate relations $Z=u\cosh \eta$ and $cT=u\sinh \eta$ we have
\begin{align*}
\frac{\partial}{\partial cT}-\frac{\partial}{\partial Z}&=e^\eta\left(\frac{1}{u}\frac{\partial}{\partial \eta}-\frac{\partial}{\partial u}\right),\\
\frac{\partial}{\partial cT}+\frac{\partial}{\partial Z}&=e^{-\eta}\left(\frac{1}{u}\frac{\partial}{\partial \eta}+\frac{\partial}{\partial u}\right),
\end{align*}
leading to
\begin{align}\label{DiracPreAcc}
0&=\left[-mc+i\hbar\left(\gamma^0\frac{\partial}{\partial cT}+\gamma^3\frac{\partial}{\partial Z}\right)\right]\psi\nonumber\\
&=\left[-umc+i\hbar e^{\eta\gamma^0\gamma^3}\left(\gamma^0\frac{\partial}{\partial \eta}+\gamma^3u\frac{\partial}{\partial u}\right)\right]\psi.
\end{align}
Eq. (\ref{DiracPreAcc}) is exactly the Dirac equation in the lab frame. It can be transformed to the accelerated frame as follows
$$
e^{\gamma^0\gamma^3\frac{\eta}{2}}\left[-umc+i\hbar \left(\gamma^0\frac{\partial}{\partial \eta}+\gamma^3\left\{u\frac{\partial}{\partial u}+\frac{1}{2}\right\}\right)\right]\psi_R=0,
$$
which can be rewritten in the more compact form
\begin{align}\label{DiracAcc}
\left[-\kappa u+i\left(\gamma^0\frac{\partial}{\partial \eta}+\gamma^3\left\{u\frac{\partial}{\partial u}+\frac{1}{2}\right\}\right)\right]\psi_R=0,
\end{align}
with $\psi_R=e^{-\gamma^0\gamma^3\frac{\eta}{2}}\psi$, where $\psi$ is the solution of the free Dirac equation in the lab frame while $\psi_R$ is the solution in the Rindler (a.k.a accelerated) reference frame.

\subsection{Constructing the superposition for a free particle}
Equipped with the spinor (\ref{RindlerSpinor2aA}) and the relationship $\psi_R=e^{-\gamma^0\gamma^3\frac{\eta}{2}}\psi$, here we will build the self-accelerating wavepacket in the lab frame.
In the Chiral representation, the Dirac spinor for a spin up electron in its rest frame with respect to a global inertial frame is 
\begin{align}\label{restSpinorA}
\psi=h(0)\left(\begin{array}{c}
1
   \\
0
   \\
1
   \\
0
\end{array}
\right)e^{-\frac{imc^2}{\hbar}T}
\end{align}
where $h(0)$ is some momentum dependent envelop function.
Let us apply a boost to a frame moving along the $Z$-axis with
momentum $p$
\begin{align}\label{boostedSpinorA}
\psi_p=h(p)\left(
\begin{array}{c}
\frac{mc-p+E_p}{2\sqrt{mc(mc+E_p)}}
   \\
0
   \\
\frac{mc+p+E_p}{2\sqrt{mc(mc+E_p)}}
   \\
0
\end{array}
\right)e^{-\frac{i}{\hbar}(E_pT-pZ)}
\end{align}
with $E_p=\sqrt{m^2c^2+p^2}$.  Now we build a wavepacket by integrating over $p$ the spinor(\ref{boostedSpinorA})
\begin{align}\label{wavepacketA}
\psi(t,z)=\int_{-\infty}^{\infty}\frac{ch(p)dp}{2E_p}\left(
\begin{array}{c}
\frac{mc-p+E_p}{2\sqrt{mc(mc+E_p)}}
   \\
0
   \\
\frac{mc+p+E_p}{2\sqrt{mc(mc+E_p)}}
   \\
0
\end{array}
\right)e^{-\frac{i}{\hbar}(E_pT-pZ)}
\end{align}
in which $\frac{cdp}{2E_p}$ renders the integral Lorentz invariant.

Let us now choose the following envelop function
\begin{align}\label{envelopA}
h(p)=\mathcal{N}e^{-\frac{a}{c\hbar}E_p},
\end{align}
where $\mathcal{N}$ is a normalization constant and $a>0$ is a constant with units of length.
Upon making the variable substitution $p=mc\sinh(b)$ and including the phase factor $e^{i\alpha b}$ with $\alpha$ being a arbitrary real number in (\ref{wavepacketA}) one ends up with the desired superposition
\begin{align}\label{kaminer2A}
\psi&=\mathcal{N}\int_{-\infty}^\infty db e^{ib\alpha}\left(
\begin{array}{c}
e^{-b/2}
   \\
0
   \\
e^{b/2}
   \\
0
\end{array}
\right)\nonumber\\
&\times e^{-i\kappa(\cosh(b)(cT-ia)-\sinh(b)Z)}.
\end{align}
Due to the particular form of $h(b)$, the $b$ integration in (\ref{kaminer2A}) can be performed exactly. In order to see this, first note that, for $a>0$
\begin{align*}
&\int_{-\infty}^{+\infty}dt e^{iy\cosh(t)+i\zeta\sinh(t)-\nu t}=i\pi e^{\frac{i\nu\pi}{2}}\left(\frac{y+\zeta}{y-\zeta}\right)^{\frac{\nu}{2}}H_\nu^{(1)}(x),\\
&\int_{-\infty}^{+\infty}dt e^{iy\cosh(t)+i\zeta\sinh(t)+\nu t}=i\pi e^{\frac{i\nu\pi}{2}}\left(\frac{y-\zeta}{y+\zeta}\right)^{\frac{\nu}{2}}H_\nu^{(1)}(x)
\end{align*}
with $x=\sqrt{y^2-\zeta^2}$. Then defining
\begin{align*}
&\kappa(ia-cT)=y,\quad \zeta=\kappa Z\\
 &x=\kappa\sqrt{(ia-cT)^2-Z^2},
\end{align*}
leads to
\begin{align}\label{kaminerandlaser3A}
\psi&=\mathcal{N}\int_{-\infty}^\infty db e^{ib\alpha}\left(
\begin{array}{c}
e^{-b/2}
   \\
0
   \\
e^{b/2}
   \\
0
\end{array}
\right) e^{iy\cosh(b)+i\zeta\sinh(b)}.
\end{align}
Before continuing, note that $ix=i\kappa\sqrt{(a+icT)^2+Z^2}$ and $H_{\nu}^{(1)}(ix)=\frac{2K_\nu(x)}{\pi i^{1+\nu}}$.
Hence, by
performing the $b$ integration one gets
\begin{align}\label{kaminerandlaser4A}
\psi&=\mathcal{N}\left(
\begin{array}{c}
F_{i\alpha-1/2}(\bar{x})
   \\
0
   \\
F_{i\alpha+1/2}(\bar{x})
   \\
0
\end{array}
\right),
\end{align}
where
\begin{align*}
\bar{a}&=\kappa a,\, \bar{Z}=\kappa Z,\, \bar{T}=\kappa cT,\\
 \bar{\omega}&=\frac{\omega}{c\kappa},\,\bar{\xi}=\bar{\omega}(\bar{T}-\bar{Z}),\, \bar{x}=\kappa x,\\
 F_{i\alpha\pm1/2}(\bar{x})&=2\left(\frac{i\bar{a}-\bar{T}-\bar{Z}}{i\bar{a}-\bar{T}+\bar{Z}}\right)^{\pm1/4+i\alpha/2}K_{\pm1/2+i\alpha}(\bar{x}).
\end{align*}
\section{Asymptotic expansions: wavepacket leaking}\label{C}
Let us begin with the spinor in the accelerated frame, which is related to (\ref{kaminerandlaser4A}) in the same way as the spinor (\ref{RindlerSpinor2aA}) is related to the one in the Lab. frame.
We have
\begin{align}\label{SpaccA}
\psi&=\mathcal{N}\left(
\begin{array}{c}
e^{ \frac{\eta}{2}}G_{i\alpha-1/2}(\bar{\zeta})
   \\
0
   \\
e^{ -\frac{\eta}{2}}G_{i\alpha+1/2}(\bar{\zeta})
   \\
0
\end{array}
\right),\nonumber\\
 G_{i\alpha\pm1/2}(\bar{\zeta})&=\left(\frac{e^\eta \left(-\bar{u} e^\eta+i \bar{a}\right)}{\bar{u}+i \bar{a} e^\eta}\right)^{\pm\frac{1}{4}+\frac{i \alpha }{2}}K_{\pm1/2+i\alpha}(\bar{\zeta}),\nonumber\\
 \bar{\zeta}&= \sqrt{\bar{a}^2+2 i \bar{a} \bar{u} \sinh (\eta)+\bar{u}^2}.
\end{align}
Now, for $\eta\gg1$ we have
\begin{align*}
&\bar{\zeta}\approx e^{i\pi/4}\sqrt{2\bar{a}\bar{u}}e^{\eta/2},\,K_{\pm1/2+i\alpha}(\bar{\zeta})\approx\sqrt{\frac{\pi}{2\bar{\zeta}}}e^{-\bar{\zeta}},\\
&\left(\frac{e^\eta \left(-\bar{u} e^\eta+i \bar{a}\right)}{\bar{u}+i \bar{a} e^\eta}\right)^{\pm\frac{1}{4}+\frac{i \alpha }{2}}\approx\left(\frac{i \bar{u} e^\eta}{\bar{a}}\right)^{\pm\frac{1}{4}+\frac{i \alpha }{2}}.
\end{align*}
Thus
\begin{align}
\psi^\dagger\psi\approx\frac{e^{-2 \left(e^{\eta}\bar{a} \bar{u}\right)^{1/2}-\frac{\pi  \alpha }{2}}}{\sqrt{2} \bar{u}}.
\end{align}
The same estimation with the Rindler spinor (\ref{PreRindlerSpinor}) leads to
\begin{align}
\psi_R^\dagger\psi_R\approx\frac{e^{-2\bar{u}+\frac{\pi \Omega }{2}}}{\sqrt{2} \bar{u}}.
\end{align}
Both asymptotic expansions will coincide if the following condition holds
\begin{align}
\frac{e^\eta\bar{a}}{\bar{u}}\sim1\Rightarrow\frac{\bar{a}}{|\bar{Z}-\bar{T}|}\sim1\Rightarrow\bar{Z}-\bar{T}\gtrsim\bar{a}.
\end{align} 
Considering that the equation for the rightmost hyperbolic trajectory as a function of time is $\bar{Z}(\bar{T})\approx\sqrt{\alpha^2+\bar{T}^2}$ (because $\bar{Z}_0\approx\alpha$ for such trajectory) we have
\begin{align}\label{avalue}
T\lesssim\frac{\hbar}{mc^2}\frac{\alpha^2-\bar{a}^2}{2\bar{a}}.
\end{align}
For instance
\begin{align*}
T\lesssim0.73\times10^{-15}s,\quad\mbox{for}\,(\bar{a},\alpha)=(0.005,30),\\
T\lesssim3.64\times10^{-15}s,\quad\mbox{for}\,(\bar{a},\alpha)=(0.001,30).
\end{align*}
\section{Wavepacket variance and norm}\label{D}
The wavepacket variance with respect to both the laboratory frame and the accelerated frame are defined as
\begin{align}
&\delta\bar{Z}=\sqrt{\langle\bar{Z}^2\rangle-\langle\bar{Z}\rangle^2},\,\langle \bar{Z}^n\rangle=\int_{-\infty}^{+\infty}d\bar{Z}\,\bar{Z}^n\psi^\dagger\psi,\\
&\delta\bar{u}=\sqrt{\langle\bar{u}^2\rangle-\langle\bar{u}\rangle^2},\,\langle \bar{u}^n\rangle=\int_{0}^{\infty}d\bar{u}\,\bar{u}^n\psi^\dagger_R\psi_R,
\end{align} 
while the wavepacket norm in both frames is
\begin{align}
&|\psi|^2=\int_{-\infty}^{+\infty}d\bar{Z}\,\psi^\dagger\psi,\,|\psi_R|^2=\int_{0}^{\infty}d\bar{u}\,\psi^\dagger_R\psi_R.
\end{align} 
For the spinor (\ref{kaminerandlaser4A}) these integrals can be calculated exactly. The results are
\begin{align}
&\langle \bar{Z}^2\rangle=K_1(2  \bar{a})\frac{ \bar{a} \left(4 (\alpha ^2-\bar{T}^2)+4 \pi   \bar{a} \pmb{L}_0(2  \bar{a}) \left(\alpha
   ^2-\bar{T}^2\right)+1\right)}{2 K_0(2  \bar{a})}\nonumber\\
   &-2 \pi   \bar{a}^2 \pmb{L}_{-1}(2  \bar{a}) (\bar{T}-\alpha ) (\alpha
   +\bar{T})+\frac{\pi   \bar{a} (\bar{T}-\alpha ) (\alpha +\bar{T})}{K_0(2  \bar{a})}\nonumber,\\
 &  +\bar{T}^2,\\
&\langle \bar{Z}\rangle=\pi \alpha\left( \bar{a} \pmb{L}_{-1}(2  \bar{a})+\frac{(2  \bar{a} \pmb{L}_0(2  \bar{a})
   K_1(2  \bar{a})-1)}{2 K_0(2  \bar{a})}\right),
\end{align}
where $K_n(2\bar{a})$ and $ \pmb{L}_{-n}(2\bar{a})$ are the Bessel and the modified Struve functions, respectively. One can extract the important result for variance $\delta\bar{Z}^2=\langle\bar{Z}^2\rangle-\langle\bar{Z}\rangle^2$:
\begin{align}\label{zspreading}
&\delta\bar{Z}(\alpha)^2-\delta\bar{Z}(0)^2=\pi\alpha ^2\Bigg[ \bar{a}^2 \pmb{L}_{-1}(2 \bar{a}) (2-\pi  \pmb{L}_{-1}(2 \bar{a}))\nonumber\\
&-\frac{\pi  (1-2 \bar{a}
   \pmb{L}_0(2 \bar{a}) K_1(2 \bar{a})){}^2}{4 K_0(2 \bar{a}){}^2}+\frac{\bar{a} (\pi  \pmb{L}_{-1}(2 \bar{a})-1)}{K_0(2 \bar{a})}\nonumber\\
   &+\frac{\bar{a} \left(\frac{2}{\pi }-2 \bar{a} (\pi  \pmb{L}_{-1}(2 \bar{a})-1) \pmb{L}_0(2
   \bar{a})\right) K_1(2 \bar{a})}{K_0(2 \bar{a})}\Bigg]
\end{align}
which is independent of time.
 \section{Decomposition of $\bar{\mathcal{R}}$ into boosts and rotations}\label{E}
In the simplified case with $a=0$ and $\alpha=0$ it is straightforward to see the following relationship
 \begin{align}
\bar{\mathcal{R}}&=e^{-\frac{\eta'\gamma^0\gamma^3}{2}}UB,\nonumber\\
 &U=e^{-\theta\left(\gamma^1\gamma^3\frac{\dot{f}_1(\xi)}{\sqrt{\dot{f}_1(\xi)^2+\dot{f}_2(\xi)^2}}+\gamma^2\gamma^3\frac{\dot{f}_2(\xi)}{\sqrt{\dot{f}_1(\xi)^2+\dot{f}_2(\xi)^2}}\right)},\nonumber\\
 &B=e^{-w(V_1\gamma^0\gamma^1+V_2\gamma^0\gamma^2+V_3\gamma^0\gamma^3)}
 \end{align}
 where
 \begin{align*}
 \theta&=\tan^{-1}\left(\frac{\sqrt{\dot{f}_1(\xi)^2+\dot{f}_2(\xi)^2}}{2mc}\right),\\
 w&=\tanh^{-1}\left(\frac{\sqrt{\dot{f}_1(\xi)^2+\dot{f}_2(\xi)^2}}{2mc\sqrt{1+\left(\frac{\sqrt{\dot{f}_1(\xi)^2+\dot{f}_2(\xi)^2}}{2mc}\right)^2}}\right)\\
 &=\tanh^{-1}(\sin\theta),\\
 V_1&=\frac{\dot{f}_1(\xi)}{\sqrt{\dot{f}_1(\xi)^2+\dot{f}_2(\xi)^2}}\cos\theta,\,V_2=\frac{\dot{f}_2(\xi)}{\sqrt{\dot{f}_1(\xi)^2+\dot{f}_2(\xi)^2}}\cos\theta,\\
 V_3&=\sin\theta.\\
 \end{align*}
 Incidentally, the boost $B$ leads to the following proper velocity

 \begin{align*}
\frac{ \mathbf{u}}{c}&=B^2\gamma_0=\gamma(\mathbf{1}+\gamma^0\gamma^k\beta_k)\\
 \gamma&=1+\frac{\dot{f}_1(\xi)^2+\dot{f}_2(\xi)^2}{2m^2c^2},\\
 \vec{\beta}&=\left(\frac{\dot{f}_1(\xi)}{mc\left(1+\frac{\dot{f}_1(\xi)^2+\dot{f}_2(\xi)^2}{2m^2c^2}\right)},\frac{\dot{f}_2(\xi)}{mc\left(1+\frac{\dot{f}_1(\xi)^2+\dot{f}_2(\xi)^2}{2m^2c^2}\right)},\frac{\dot{f}_1(\xi)^2+\dot{f}_2(\xi)^2}{2m^2c^2\left(1+\frac{\dot{f}_1(\xi)^2+\dot{f}_2(\xi)^2}{2m^2c^2}\right)}\right).
 \end{align*}

This is expected since the solutions to the classical and quantum equations of motion for an electron in a laser field are, up to the phase factor to the right of the matrix spinor (i.e., $\Psi$ on the main text), the same.

\section{Experimental feasibility}\label{F}

\subsection{Creation of nonspreading wavepackets}

Our findings have a direct implication for the experimental realization of a laser-driven collider.  
We assume to use the version of a laser-driven collider based on the setup of a high-energy gamma-photon beam counterpropagating an ultrastrong laser field. The electrons and positrons are created inside the laser field due to the nonlinear Breit-Wheeler process. They are accelerated by the laser field within a cycle of the laser field, and collide, initiating a high-energy electron-positron collision reaction (see Ref. [10] of the paper). Two questions should be addressed: can the nonspreading feature of the created electron and positron wavepacket be designed and is the recollision time short enough to maintain the nonspreading character of the wavepacket?

To answer these questions, let us estimate the velocity (or the Lorentz $\gamma$-factor) of the average rest frame (RF) of the created pair at the threshold of the process for the given gamma-photon energy $\Omega_0$ in the Lab-frame (LF), and the laser field strength $a_0$. Here $a_0\equiv eE_0/m\omega$ is the classical strong field parameter of the laser field, where the quasimomentum (momentum averaged over the laser period) of the electron and positron is vanishing, $\mathbf{q}=0$, $q_0=m_*$ and where $m_*=m\sqrt(1+a_0^2/2)$ is the dressed mass of the electron in a linearly polarized laser field. As the pair is created by absorbing one gamma-photon of the energy $\Omega_0'$  (in RF) and $n$ counterpropagating laser photons with an energy $\Omega_0'$ (in RF), the energy-momentum conservation law in RF at the threshold of the process yields: $\Omega_0'=n\omega'=m_*$. In an ultrastrong laser field $a_0\gg 1$, the average number of laser photons involved in the pair production process is $n\sim a_0^3$. We choose the gamma-photon energy in LF to fulfill the condition $\Omega_0 >  a_0^3\omega$. In this case the RF propagates along the gamma-photon and the RF's $\gamma$-factor is determined from the Doppler-shifted momentum conservation condition: $\Omega_0/(2\gamma)=2\gamma n\omega\approx ma_0/\sqrt{2}$. Thus, with the given $a_0$, the RF's $\gamma$ of the most probable pair production is determined from the condition
\begin{equation}\label{gamma}
 2\sqrt{2}a_0^2\gamma \omega/m=1,
\end{equation}
which will require the gamma-photon energy
\begin{equation}\label{Omega}
 \Omega_0\approx \sqrt{2}m\gamma a_0.
\end{equation}
Assuming an infrared laser field with $\omega/m=10^{-6}$,   $a_0=10^2$ (the laser intensity of $10^{22}$~W/cm$^2$), we have from Eqs.~(\ref{gamma})-(\ref{Omega}) $\gamma\approx 30$, and $\Omega_0\approx 2$ GeV.

The recollision of the created pair takes place after the excursion of the electron and positron during the period of the laser field. As the RF moves along the gamma-photon, i.e., opposite to the laser wave propagation direction, in RF the laser frequency is up-shifted $\omega'=2\omega\gamma$, and the recollision time in LF is $T'=T/\gamma\approx 3\times 10^{-17}$~s. From  Eq.~(14) of the revised manuscript the leaking time of the nonspreading wavepacket is $mT'_l=(\alpha^2-\overline{a}^2)/(2\overline{a})$. In the case of the optimal parameters $\alpha=30$, $\overline{a}=10^{-2}$, $T'_l=(\lambda_C/c)\alpha^2/(2\overline{a})\approx 10^{-15}$~s. Thus, for the chosen parameters the recollision time in RF is much smaller than the leaking time of the nonspreading wavepacket.

The next point is how to create the nonspreading wavepacket of Eq.~(7) of the manuscript. The essential point of this wavepacket is the specially tailored momentum chirping of the wavepacket given by the phase $\varphi(p)=\alpha b=\alpha \sinh^{-1} (p/m)$. This chirping induces a spatial shift of each momentum component in the laser field $\delta x (p)=\partial \varphi(p)/\partial p$. The created wavepacket of the electron and positron will be chirped if the particle with the corresponding momentum value is created with the corresponding spatial delay $\delta x (p)$. The particle in LF moves with the momentum $p=m\gamma$. From Eqs.~(\ref{gamma})-(\ref{Omega}), $\gamma$ is determined either by the laser field intensity $a_0^2$, or by the gamma-photon energy. Tailoring specifically the laser intensity in space according to the function $\delta x (p)$, one can achieve chirping of the created wavepackets of the electron and positron. Another possibility is to use a chirped gamma-photon beam, and in this way transfer the chirp from the gamma-photons to the created wavepackets of electrons and positrons.

\subsection{Role of radiation reaction}

The approach based on the Dirac equation can be valid if the radiation reaction does not disturb much the electron dynamics. We can formulate it as a restriction on the laser and electron parameters. The condition for negligible radiation reaction can be formulated as the radiation energy loss ($\Delta\varepsilon$) being negligibly small compared with the electron energy ($ \varepsilon$): $\Delta\varepsilon\ll \varepsilon$. In the laser-driven collider (Refs. [8,10]) the electron acceleration takes place during the excursion in a half-cycle of the laser field. As the radiation formation length is $a_0$-times smaller than the electron trajectory period at $a_0\gg 1$ (see e.g. Ref.~[6]), the number of the radiation formation lengths during one laser period is  $a_0$. Here, $a_0=eE_0/(m\omega)$ is the strong field parameter of the laser field, with the laser field amplitude $E_0$, and the frequency $\omega$. As the probability for a photon emission on a formation length is of the order of the fine-structure constant $\alpha_f$, $\Delta\varepsilon\sim \alpha_f a_0 \omega_c$, with the characteristic energy of the emitted photon $\omega_c\sim \chi \varepsilon$, where the quantum strong field parameter  $\chi\equiv E'/E_{cr}$ describes the photon recoil (see e.g. Ref.~[6]). Here, $E'$ is the background field strength in the rest frame of the electron, and $E_{cr}$ is the Schwinger critical field. In the laser collider setup, one can  estimate  $\chi\sim 2\gamma_0 (\omega/m) a_0$), when the gamma-photon with an energy $m\gamma_0$ counterpropagates the laser field as in Ref.~[10]. Thus, the condition to neglect radiation reaction in the laser collider will read $\alpha a_0 \chi \sim 2\alpha_f a_0^2\gamma_0 (\omega/m)\ll 1 $. For instance, in the case of an infrared laser field $\omega/m\sim 10^{-6}$,  GeV gamma photon $\gamma_0\sim 10^{3}$, and an ultrastrong laser field of the intensity of $10^{22}$ W/cm$^2$ ($a_0\sim 10^2$), this condition can be fulfilled $\Delta\varepsilon/\varepsilon \sim 10^{-1}$. Thus, the approach based on the Dirac equation still can be valid for typical parameters of a laser-driven collider.

\end{widetext}

\bibliography{strong_fields_bibliography}

\end{document}